\documentclass[aip,rsi,amsmath,amssymb,reprint,a4,times]{revtex4-1}
\usepackage{graphicx}
\usepackage{color}
\usepackage[utf8x]{inputenc}

\begin{document}

\title{An optically detected magnetic resonance spectrometer with tunable laser excitation and wavelength resolved infrared detection}

\author{M. Negyedi}
\affiliation{Department of Physics, Budapest University of Technology and Economics and MTA-BME Lend\"{u}let
Spintronics Research Group (PROSPIN), PoBox 91, H-1521 Budapest, Hungary}
\affiliation{Universit\"{a}t T\"{u}bingen Physikalische Institut, Auf der Morgenstelle 14D 72076 T\"{u}bingen, Germany}
\altaffiliation[Permament address: ]{HighFinesse GmbH, Auf der Morgenstelle 14D 72076 T\"{u}bingen, Germany}

\author{J. Palot\'{a}s}
\affiliation{Department of Physics, Budapest University of Technology and Economics and MTA-BME Lend\"{u}let
Spintronics Research Group (PROSPIN), PoBox 91, H-1521 Budapest, Hungary}

\author{B. Gy\"{u}re}
\affiliation{Department of Physics, Budapest University of Technology and Economics and MTA-BME Lend\"{u}let
Spintronics Research Group (PROSPIN), PoBox 91, H-1521 Budapest, Hungary}

\author{S. Dzsaber}
\affiliation{Department of Physics, Budapest University of Technology and Economics and MTA-BME Lend\"{u}let
Spintronics Research Group (PROSPIN), PoBox 91, H-1521 Budapest, Hungary}
\altaffiliation[Present address: ]{Institute of Solid State Physics, Vienna University of Technology, Vienna, Austria}

\author{S. Kollarics}
\affiliation{Department of Physics, Budapest University of Technology and Economics and MTA-BME Lend\"{u}let
Spintronics Research Group (PROSPIN), PoBox 91, H-1521 Budapest, Hungary}

\author{P. Rohringer}
\affiliation{University of Vienna, Faculty of Physics, Strudlhofgasse 4., Vienna, A-1090, Austria}

\author{T. Pichler}
\affiliation{University of Vienna, Faculty of Physics, Strudlhofgasse 4., Vienna, A-1090, Austria}

\author{F. Simon}
\email[Corresponding author: ]{f.simon@eik.bme.hu}
\affiliation{Department of Physics, Budapest University of Technology and Economics and MTA-BME Lend\"{u}let
Spintronics Research Group (PROSPIN), PoBox 91, H-1521 Budapest, Hungary}
\affiliation{University of Vienna, Faculty of Physics, Strudlhofgasse 4., Vienna, A-1090, Austria}

\date{\today}

\begin{abstract}
We present the development and performance of an optically detected magnetic resonance (ODMR) spectrometer. The spectrometer represents advances over similar instruments in three areas: i) the exciting light is a tunable laser source which covers much of the visible light range, ii) the optical signal is analyzed with a spectrograph, iii) the emitted light is detected in the near-infrared domain. The need to perform ODMR experiments on single-walled carbon nanotubes motivated the present development and we demonstrate the utility of the spectrometer on this material. The performance of the spectrometer is critically compared to similar instruments. The present development opens the way to perform ODMR studies on various new materials such as molecules and luminescent quantum dots where the emission is in the near-infrared range and requires a well-defined excitation wavelength and analysis of the scattered light.
\end{abstract}
\maketitle

\section{Introduction}
Optically detected magnetic resonance (ODMR) \cite{clarke1982triplet} combines the advantages of both magnetic resonance (high energy resolution and spin specificity) and optical methods (high sensitivity). ODMR is used extensively in various branches of research, such as e.g. studying light emitting diodes \cite{ShinarOLEDReview}, quantum computing architectures \cite{DiamondQubit}, spintronics candidate materials \cite{WrachtrupNJP}, and biomedical sciences\cite{ODMRReviewBiology}. Depending on the physical process which is being exploited, various versions of ODMR are known, such as e.g. PL- (photoluminescence), F- (fluorescence), Ph- (phosphorescence), or PA- (photoinduced absorption\cite{ShinarPRL2005}) DMR. Of these techniques the most frequently used is PLDMR which was successfully employed to detect ODMR on single molecules \cite{KohlerNature,WrachtrupNature}.

PLDMR exploits that the optical excitation creates an electron-hole pair which forms a bound quasi-particle, with 4 possible spin states (1 singlet and 3 triplet). The solid-state analogue of the electron-hole pair is the so-called \emph{exciton} which is encountered in semiconductors, macromolecules, and molecular solids. In the following, we refer to the excited electron-hole pair state as exciton. Spin-conservation only allows the optical excitation of spin singlet states, however it may cross over to a triplet state with a low probability, a spin-forbidden process known as intersystem crossing, which is the result of spin-orbit coupling. Again, spin-conservation of the radiative electron-hole pair recombination (the process known as phosphorescence) dictates that the decay rate of the triplet state to the ground state is much smaller than that of the usual fluorescence process, therefore molecules in the triplet state become "dark", i.e. they do not contribute to the photoluminescence. Both the intersystem crossing and the phosphorescence processes are driven by non spin-conserving interactions for which the leading contribution is from spin-orbit coupling. 

The three sub-levels of the triplet exciton have wavefunctions with differing orbital shape and thus differing strength of spin-orbit coupling. This results in a different population of the sub-levels when formed and a different decay rate to the ground state. The ODMR method is based on altering the populations of the triplet sub-levels thus altering the phosphorescence signal. This is achieved by irradiating the three sub-levels by microwaves under electron-spin resonance conditions; these are split either due to zero-field splitting or by an applied external magnetic field. Interestingly, it is not only the phosphorescence signal which changes upon the microwave irradiation but also the fluorescence: molecules which are otherwise trapped in long-living triplet states are \emph{liberated} and can therefore contribute to fluorescence. Detection of the ODMR provides a plethora of spectroscopic information including zero-field splitting energy values, magnitude of the singlet-triplet energy gap, optical decay rates, and spin-relaxation times.

As a result of the principle of operation, the ingredients of an ODMR spectrometer are \cite{clarke1982triplet,ODMRReviewBiology,ShinarOLEDReview}: exciting light, a light collection optics, and microwave irradiation. The commonly used method is to continuously \emph{chop} the microwave irradiation and to detect the change in the collected light in-phase with the chopping using a lock-in amplifier. Most ODMR spectrometers operate with a single-laser excitation. The excitation is filtered from the collected light and the latter is detected with a photodiode. This setup is appropriate for most studies where the molecular absorption and emission bandwidths are large and lie in the visible spectral range. 

Single-walled carbon nanotubes \cite{IijimaNAT1993} are known for their unique structural and optical properties: their one-dimensional structure results in strong and well-defined optical transitions \cite{OconnellSCI}, which are defined by the so-called folding vector index $(n,m)$ (Ref. \onlinecite{ReichBook}). The optical excitations form strongly bound excitons\cite{HeinzSCI2005} (with binding energies as high as 1 eV) for SWCNTs which are semiconducting. Photoluminescence characterization of SWCNTs has become a common tool \cite{Bachilo:Science298:2361:(2002)} and use of a tunable light source and spectrally resolving the emitted light allows to construct the so-called photoluminescence- or PL-maps. 
ODMR was studied in SWCNTs in Ref. \onlinecite{HertelNatPhot} (work published while our development was underway) but without spectral resolution or using several excitation wavelengths. This limits the ODMR information due to the simultaneous observation of several $(n,m)$ species. 

An ODMR study on molecules with well-defined optical transitions clearly requires a tunable and powerful light source, spectral resolution of the emitted light, and specifically for SWCNT, a near infrared (NIR) detection. These requirements pose several challenges for the construction: i) a powerful enough tunable light source could be a dye or Ti:Sapphire laser whose handling is cumbersome and are known for an unstable output intensity \cite{duarte1990book}, ii) light collection has to be efficient: spectrally resolving the light and measurements on a heterogeneous sample made of molecules with differing optical transitions decreases the number of available photons, iii) NIR detection is known to be more difficult due to the lower optical  sensitivity of such detectors. 

Herein, we present the development of an ODMR spectrometer which is suited to study molecules with well-defined emissions in the near-infrared. We believe that the optimal conditions for such requirements have been realized concerning spectral sensitivity, energy resolution, and the exciting laser power. The spectrometer uses a tunable laser as excitation source, the collected light is analyzed spectrally and is detected with an InGaAs detector. We discuss in detail the spectrometer setup and its performance is presented for SWCNTs, for which we present the first spectrally resolved ODMR results.

\section{The ODMR spectrometer}
\subsection{The spectrometer setup}

\begin{figure}[!htp]
\includegraphics[scale=0.48]{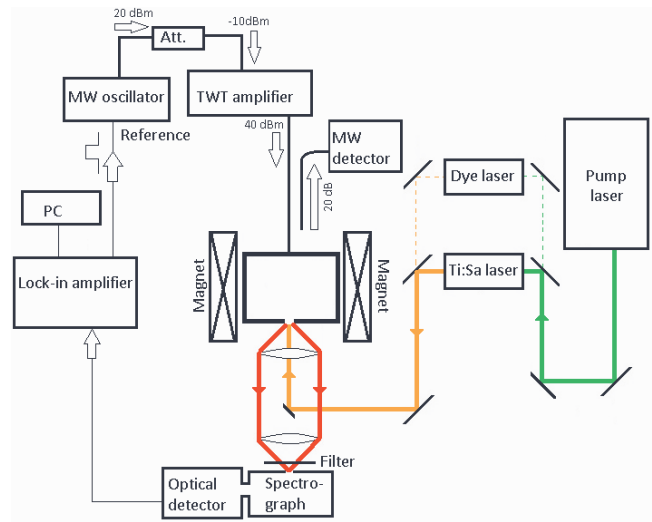}
\caption{The layout of the ODMR spectrometer. The colors of green, orange, and red represent the 532 nm pump, the 560-900 nm exciting laser, and the 1000-1500 nm scattered light, respectively. Arrows show the direction of the light propagation.}
\label{Fig1_ODMR_setup}
\end{figure}

As discussed above, ODMR spectroscopy involves the in-phase (or lock-in) detection of a scattered light from the sample while the latter is placed in a microwave cavity where the microwave intensity is modulated. Our spectrometer is different from usual ones in three respects: the exciting light is wavelength selected, the back scattered light is analyzed according to the wavelength, and the latter is in the infrared region. This constitutes challenges and results in a complex spectrometer setup as described in the following. The block diagram is shown in Fig. \ref{Fig1_ODMR_setup}., and the optical and microwave units are discussed separately below.

\subsubsection{Optical setup}
The light source of the spectrometer are two tunable lasers: a Ti:Sapphire and a dye-laser system (Radiant Dyes Laser \& Acc. GmbH, Germany) with a spectral purity of $1\,\text{cm}^{-1}$. Both are pumped by a frequency doubled Nd:YAG pump laser (Coherent Verdi 5G). The desired light source is selected by mirrors as Fig. \ref{Fig1_ODMR_setup}. hints, and the exciting beam diameter is about 2 mm. A series of dyes (Rhodamin 101, Rhodamin 6G, DCM Special, Pyridin 1-2) allows to cover the 560-700 nm range \cite{duarte1990book} and the Ti:Sapphire laser functions for the 700-900 nm range with typical powers of 50-300 mW. 

The light from the tunable lasers is guided by highly reflecting dielectric coated mirrors to a small (2 mm) right angle prism mirror which is followed by an achromatic doublet lens (Thorlabs Inc. AC254-030-B, $d$=25.4 mm $f$=30 mm) with an antireflection coating for 650-1050 nm. The lens functions as both excitation focusing and scattered light collection and has a high collection efficiency with f/\#=1.2. We found that the achromatic doublet lens represents a simple, yet effective construction (with low spherical and chromatic aberration) and it has a small size as compared to a more involved lens system \cite{smith2007modern}. The focus diameter on the sample is about 20 microns. Although the coating is not optimal for either the excitation or collection, it is an optimal compromise for both ranges. The backscattering or 180 degree geometry of light collection, is proper in our case as it provides an easy adjustment and side access for the microwave cavity. 

The collected light forms a collimated beam with a diameter of about 25 mm (indicated by two red lines in Fig. \ref{Fig1_ODMR_setup}.) which is focused on the entrance slit of the spectrograph by another achromatic doublet lens (Thorlabs Inc. AC254-100-B, $d$=25.4 mm $f$=100 mm) whose f/\# matches that of the spectrograph (f/\#=4.1). The two lens setup is known as an "image relay" and it provides an image size on the entrance slit magnified by the ratio of the two focus distances which gives an image diameter of 100 microns in our case. We use a long pass filter (Thorlabs FEL0900, edge at 900 nm, optical density of 6) before the entrance slit to prevent the exciting light from entering into the spectrometer. 

The Czerny-Turner spectrograph (Horiba JY iHR320) has a single grating (600 gr/mm, blazing wavelength 1000 nm) and a focal length of 320 mm and is equipped with a single-channel liquid nitrogen cooled InGaAs detector (Electro-Optical Systems Inc., IGA1.9010L, 1 mm active size, Noise Equivalent Power/NEP=$10^{-14}\,\text{W}/\sqrt{\text{Hz}}$, electronic bandwidth/6 dB point is 2.2 kHz) which is optimized for 1000-2000 nm wavelength range. The spectrograph has a resolution of 0.5 nm (or $5\,\text{cm}^{-1}$ at 1000 nm) for an entrance slit larger than 100 microns, which is sufficient for most applications (HWHM of the SWCNT photoluminescent peaks is typically 20 nm). Larger slit settings do not reduce the resolution as it is essentially set by the image size on the entrance slit. The detector current is converted to voltage and is measured simultaneously with a DC voltmeter, which provides the PL signal, and with a lock-in amplifier (Standford Research Systems, SR830), which provides the ODMR signal.

\subsubsection{Magnetic resonance setup}
The source of the microwave signal is a HP83751B signal generator (output: 2-20 GHz). This source is used at maximum output power (20 dBm or 100 mW) as otherwise the output has a sizeable amplitude noise. This is passively attenuated to -10 dBm (0.1 mW) and is fed into a TWT (travelling wave tube) amplifier (Varian WZX6980G2, 8-12 GHz, gain of 50 dB, saturated power 40 dBm) which provides a microwave output power of 40 dBm (10 W). The microwave signal is chopped by a square wave from the lock-in amplifier at typically 1 kHz frequency. This microwave signal is fed into a home-built TE011 cylindrical microwave cavity whose resonance frequency is around 10 GHz (Ref. \onlinecite{PooleBook}) and is critically coupled. A microwave detector on the coupled arm (20 dB) of a directional coupler (directivity $>40\,\text{dB}$) monitors the power reflected from the cavity and it also serves to frequency lock the source to the cavity resonance using a home-built automatic frequency control (AFC) circuit \cite{PooleBook}. The DC magnetic field is provided by an electromagnet stabilized by a Hall-probe and a feedback system.


\subsection{Measurement considerations}

Besides selecting the exciting laser and analyzing the energy of the scattered light, our instrument follows the setup of the conventional ODMR spectrometers \cite{clarke1982triplet,ShinarPRB1996}: the exciting light is unmodulated and the scattered light is detected in-phase with the chopped microwaves. However, the detection of near-infrared scattered light imposes some constraints: NIR detectors are known to be less sensitive than their visible counterparts (due to the lower photon energy) which results in a trade-off in the detector band-width. Most commercially available high sensitivity NIR detectors have a band-width in the 100 Hz-1 kHz range. This means that the microwave chopping frequency is limited to the same range, too. 

Selection of the optimal frequency for the measurement relies on the frequency characteristics of the noise sources. We identified the following noise sources of our system: intensity fluctuations of the exciting laser (or oscillator noise), mechanical instabilities due to sample vibration, detector noise, and shot noise. All noise sources have to be considered in view of the expected ODMR signal. The typical ODMR signal intensity versus the photoluminescence intensity is $\Delta \text{PL}/\text{PL}\approx 10^{-5}..10^{-6}$. 

Of the above noise sources, the mechanical noise could be reduced well below the other contributions. The detector noise has a white frequency characteristics with the above given NEP for the 10-1000 Hz range with a 6 dB point at 2200 Hz due to the low frequency pass detector filtering. Lasers have inherent intensity fluctuations due to the presence of spontaneous emission which competes with stimulated emission. This is characterized by the \emph{relative intensity noise} (RIN), which describes the noise power per unit frequency bandwidth relative to the output power at a specified frequency separation from the carrier. The oscillator noise is usually strongly frequency dependent. We measured the RIN for our tunable lasers using a fast photodiode connected to a lock-in amplifier and obtained that RIN follows roughly $\text{RIN}(f)=-30 -8\cdot \text{log}_{10}(f)\,\text{[dBc/}{\sqrt{\text{Hz}}}]$ up to 1 kHz, above which the low-pass filtering limits the measurement. It means that e.g. $\text{RIN}=30\,\text{[dBc/}{\sqrt{\text{Hz}}}]$ at $1\,\text{Hz}$ frequency and $\text{RIN}=-54\,\text{[dBc/}{\sqrt{\text{Hz}}}]$ at $1\,\text{kHz}$. The latter value means that for 1 kHz microwave chopping frequency, our instrument can detect ODMR signals as low as $\Delta \text{PL}/\text{PL}=4\cdot 10^{-6}$ for a time constant of 1 second. We note that a commercially available \emph{noise eater} did not substantially reduce the oscillator noise in our operating frequency range.

The shot-noise has a white frequency characteristics and depends solely on the power of the detected light. This in turn depends on the photoluminescence intensity of the investigated signal. The detector sensitivity limit of $10^{-14}\,\text{W}$ corresponds to $N_{\text{NEP}}=6\cdot 10^{4}$ photons per seconds for 1 eV photons ($\lambda=1240\,\text{nm}$). Our NIR detector electronics saturates for an incoming power of $10^{-9}\,\text{W}$, which corresponds to an incoming photon flux of $N_{\text{satur}}=6\cdot 10^{9}$, which gives a shot-noise that is 25 \% larger than the detector noise. It means that the incoming PL intensity is optimal if it nearly saturates the detector electronics. In this case, the three contributions to the noise (detector noise, oscillator noise, shot-noise) have the same order of magnitude. We found that this near-saturation occurs for the investigated carbon nanotube suspensions if the exciting laser power was about 200 mW. This value however depends on the sample concentration and whether the resonance of the photoluminensce excitation is achieved. In addition, other effects such as sample heating has to be kept in mind in order to select the optimal exciting laser power.

Summarizing the optimal conditions, we find that a 1 kHz microwave chopping frequency and a laser power as high as to nearly avoid saturation of the NIR detector (or sample heating) should be used.

\section{Performance of the spectrometer on single-walled carbon nanotubes}

\subsection{Sample preparation and sample holder}
Single-walled carbon nanotube samples, manufactured by the \emph{High-pressure Carbon Monoxide method} or \emph{HiPco} (Carbon Nanotechnologies Inc.) were used. The diameter distribution in such samples follows a Gaussian with a mean and variance of 1 nm and 0.1 nm, respectively. It means that more than 100 different SWCNTs (with differing $(n,m)$ indices) are present in such samples with varying abundance. These SWCNTs have all different absorption and emission photoluminescence (PL) transition energies and PL is the standard method to characterize SWCNT abundance \cite{Bachilo:Science298:2361:(2002),WeismanNL2003}. 

We followed the standard procedure to produce photoluminescing SWCNTs \cite{OconnellSCI,HeinzPRL2004}: the nanotubes were introduced into a 2 \% solution of DOC (Sodium deoxycholate) in distilled water with a concentration of about 1 mg/L. This mixture was tip-sonicated for about 5~hours, which was followed by ultracentrifugation at 400,000~g's for half an hour. This suspension contains mainly individual nanotubes which are wrapped by the surfactant molecules \cite{Bachilo:Science298:2361:(2002),WeismanNL2003} with an unknown concentration. The samples were placed in quartz ampoules with 3 mm outer diameter, which were sealed hermetically. We used a double-walled quartz cryostat filled with liquid nitrogen to perform measurements at a 77~K sample temperature. Measurements are only possible for frozen samples due to the substantial microwave absorption of water. Allowing for variable temperature ODMR measurements in our setup is the subject of future developments. 

\subsection{Spectrometer sensitivity}

\begin{figure}[!htp]
\includegraphics[scale=0.45]{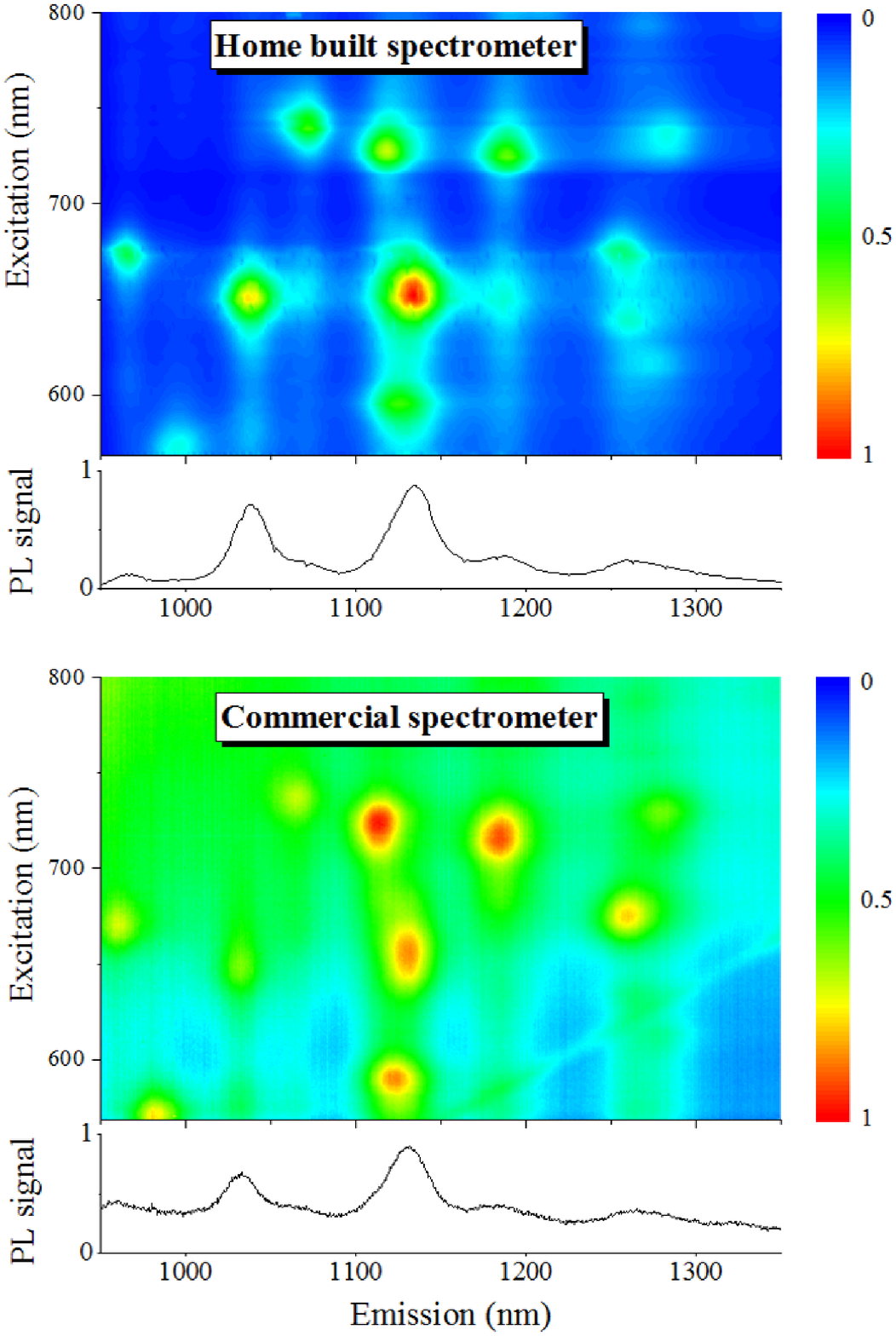}
\caption{Comparison of the PL-maps of SWCNTs measured with a commercial and the home-built spectrometer. Two corresponding individual PL spectra are shown for a 650 nm excitation. Note the smaller contrast and a significant background in the top left corner for the measurement with the commercial spectrometer.}
\label{Fig2_PL_commercial_homebuilt}
\end{figure}

As a first step, we compare photoluminescence measurements on SWCNTs performed with a commercial PL spectrometer (Horiba Jobin Yvon, Fluorolog) and our instrument in order to assess its overall optical sensitivity. The result is shown in Fig. \ref{Fig2_PL_commercial_homebuilt}. in the form of a so-called PL-map and two individual PL spectra. The earlier 3D contour plot presentation is obtained by combining the individual PL spectra. The commercial spectrometer uses a 450 W Xenon lamp, an f/\#=4 collection optics and a liquid nitrogen cooled 1024 pixel InGaAs detector. The lamp output is filtered with a double monochromator which yields 1 mW power for a 1 nm bandwidth around 600 nm excitation. This bandwidth or spectral purity is sufficient concerning the linewidth of excitation profile of the PL spectra of SWCNTs and its further reduction would reduce the power further. The amount of collected photons goes with the square of the f/\# and the number of pixels improves the signal-to-noise ratio by $\sqrt{1024}$ for a given integration time. Altogether, our spectrometer (f/\#=1.2, 100 mW excitation, single channel monochromator) is expected to possess a factor 30 higher sensitivity for the same measurement time. This is supported by Fig. \ref{Fig2_PL_commercial_homebuilt}. where a factor 40 times larger signal-to-noise ratio was observed for the same measurement time.

The home-built spectrometer has two additional advantages: the filtering of the excitation is more efficient due to the use of edge filters rather than a simple colored filter in the commercial one. This results in the absence of significant stray light as seen in the top left corner of Fig. \ref{Fig2_PL_commercial_homebuilt}. Another advantage is the absence of a baseline as the single-channel detector has a low dark current. The InGaAs array detector has a relatively large dark current and a pixel dependent sensitivity. This results in a pixel dependent background which is usually removed by measuring the dark background for some time and subtracting it from the real measurement. In fact the ODMR measurement cannot be performed for an array detector with ease and it would require a so-called gated CCD operation thus the single-channel detection is better suited in our case. A disadvantage of the present spectrometer is that laser excitation wavelength change is not automatized and even if it could be automatized for a narrower (maximum 20 nm) excitation range, user intervention is inevitable given that a dye change or retuning the laser resonator over the full operational range is required.

\subsection{ODMR measurements on SWCNTs}

\begin{figure}[!htp]
\begin{center}
\includegraphics[scale=0.4]{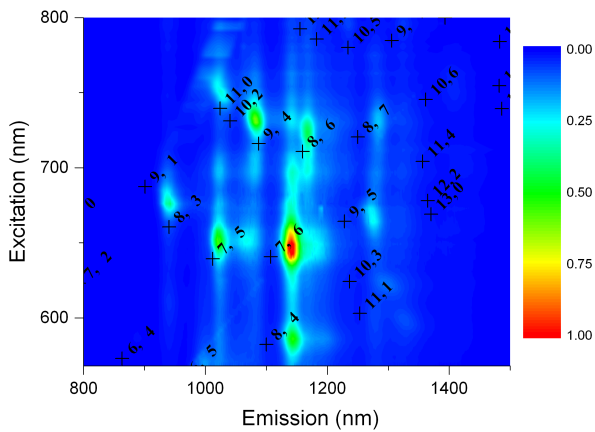}
\caption{PL-map of SWCNTs at 77 K. The crosses show the PL peaks from Ref. \onlinecite{WeismanNL2003} along with the $(n,m)$ chirality indices. The PL peak positions differ from those in the present study due to the use of different surfactant molecules.}
\label{Fig3_PL_map_77K}
\end{center}
\end{figure}

In principle, our setup allows to obtain the ODMR data with respect to three variables: the magnitude of the DC magnetic field, the excitation, and the emission wavelength. At the time of performing our measurements, the first report on ODMR in SWCNTs in Ref. \onlinecite{HertelNatPhot} was not available. If the ODMR signal exists at all, it is expected to be observable with the highest probability for a strong PL transition. In Fig. \ref{Fig3_PL_map_77K}, we show PL-map at 77 K for the SWCNT sample. We label the PL peaks with the $(n,m)$ indices from Ref. \onlinecite{WeismanNL2003} for comparison. The latter data was obtained at 300 K for SWCNTs wrapped in SDS (Sodium dodecyl sulfate) surfactant molecules. The use of different surfactant in our study (DOC) explains the difference in the PL peaks between the literature and our study. The PL peak positions shift further upon cooling the sample to 77 K, we thus took the PL excitation and emission energies from the PL-map at 77 K.

\begin{figure}[!htp]
\includegraphics[scale=0.45]{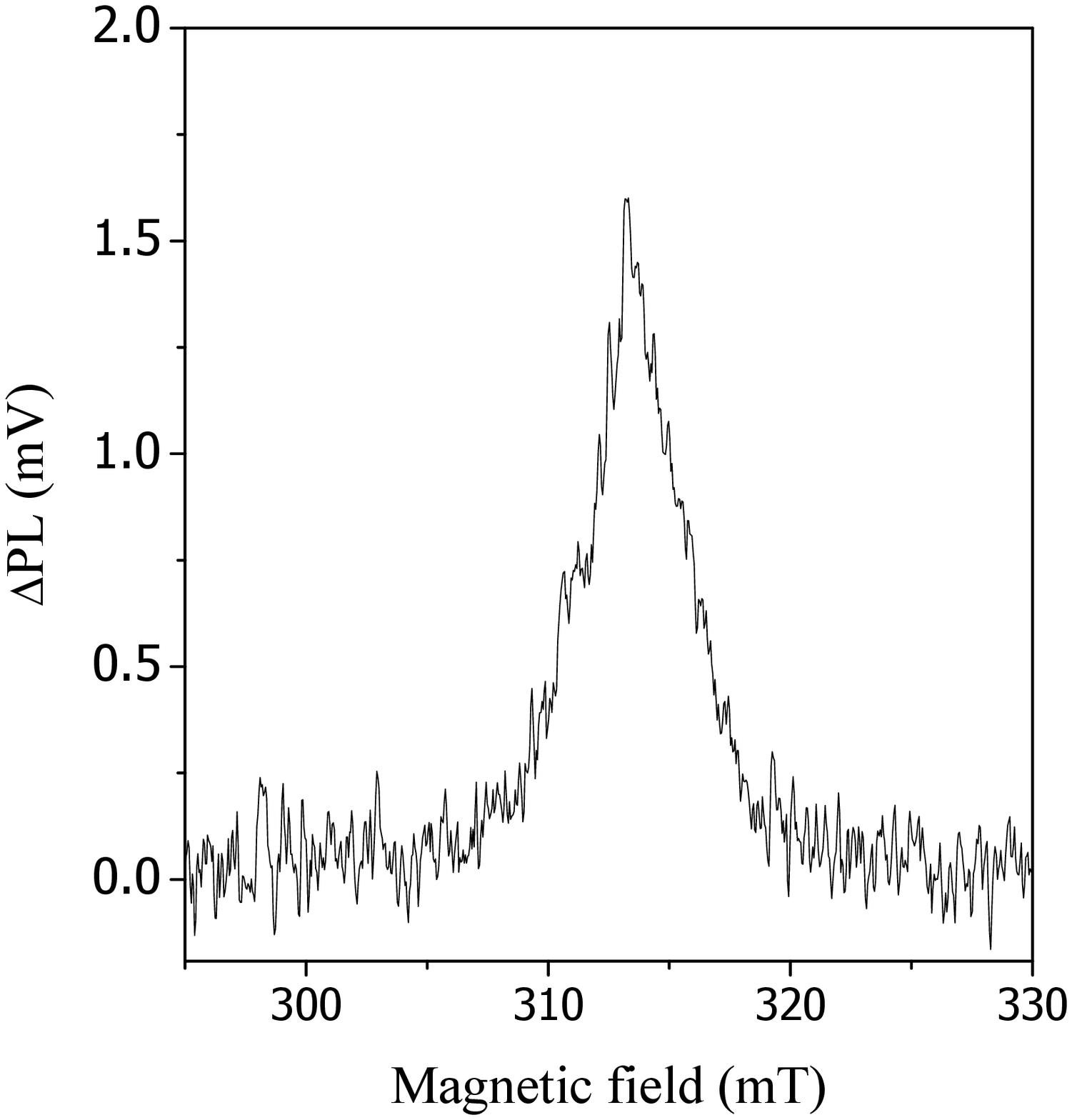}
\caption{Magnetic field swept ODMR signal. Note the non-Lorentzian lineshape of the data.}
\label{Fig4_field_swept_ODMR}
\end{figure}

We chose the $\lambda_{\text{excitation}}=730\,\text{nm}$ and $\lambda_{\text{emission}}=1086\,\text{nm}$ PL peak and performed a magnetic field sweep while observing the ODMR signal with the lock-in amplifier. The result is shown in Fig. \ref{Fig4_field_swept_ODMR}. We show the ODMR signal, or $\Delta\text{PL}$, as a detected voltage for clarity: we show below that the ratio of the PL and ODMR signals depends on the excitation and emission wavelengths. Therefore showing the ODMR signal in units relative to the PL (as it is customary in the ODMR literature) would be misleading.

Clearly, a peak is observed in the ODMR signal as a function of the magnetic field around a $g$-factor value of 2 and an FWHM of about 5 mT. The lineshape is non-Lorentzian. The spectral details of the line (position, linewidth, lineshape, and dependence on $(n,m)$) are discussed elsewhere and we focus on the ODMR spectrometer performance. We fixed the magnetic field to the maximum of the field swept ODMR signal in order to perform a wavelength resolved ODMR study. In principle, our multidimensional mapping capability raises the possibility for a number of interesting experiments, such as e.g. studying the ODMR signal while keeping the magnetic field on the wings of the non-Lorentzian signal. These studies are, however, beyond the scope of the present work. 

\begin{figure}[!htp]
\includegraphics[scale=0.45]{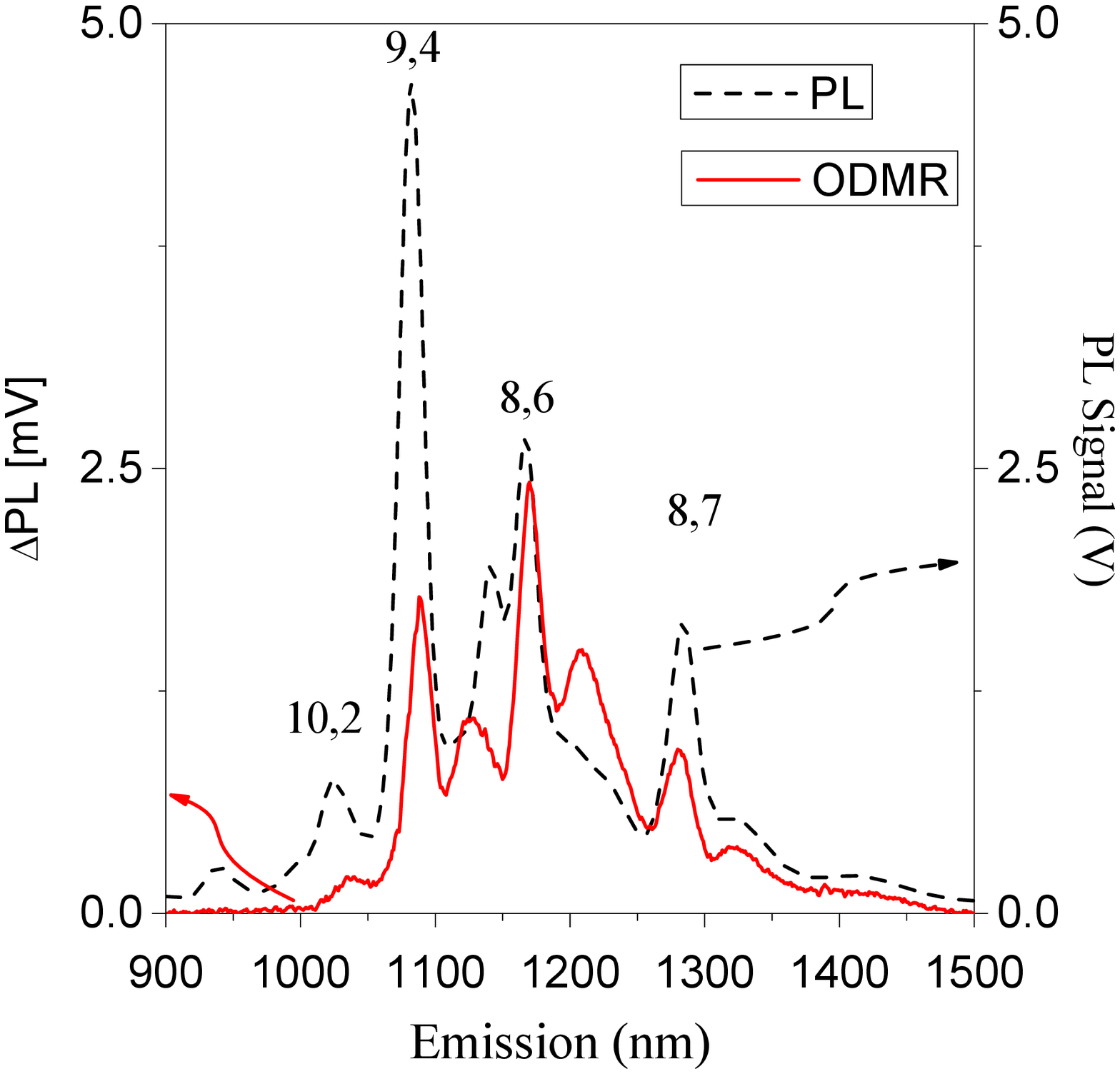}
\caption{The simultaneously measured PL (dashed curve) and $\Delta\text{PL}$ (or ODMR) signals of SWCNTs for 730 nm excitation. The ODMR signal (solid curve) is shown on a 1000 times enlarged scale. The $(n,m)$ SWCNT indices are also shown. Note that the relative intensity of the PL and ODMR signals is not constant in the spectra.}
\label{Fig5_ODMR_PL_compare}
\end{figure}

In Fig. \ref{Fig5_ODMR_PL_compare}. we show the simultaneously detected PL and ODMR spectra for 730 nm excitation. The spectrum scan time was 1 minute, which involved 600 spectrum points, i.e. 100 ms/point. We observe changes in the PL spectrum in-phase with the microwave chopping, $\Delta\text{PL}$, i.e. an ODMR signal which is approximately 1000 times smaller than the PL signal itself. However, the relative intensity of the PL and $\Delta\text{PL}$ is not constant throughout the spectra, which requires further investigation. We find that the ODMR signal has a signal-to-noise ratio of about 100, due to the fact that we designed our instrument to reach a $\Delta\text{PL}/\text{PL}$ sensitivity of $10^{5}$. This means that our instrument is indeed capable of detecting the small light intensity variations which are associated with the ODMR technique in a spectrally resolved manner in the near-infrared.

\section{Conclusions}

In conclusion, we presented the development and the performance characterization of an ODMR spectrometer. It represents advances in three parameters: a tunable laser excitation, a wavelength resolved scattered light detection and operation in the near-infrared range. It is ideally suited for the ODMR studies of macromolecules or quantum dots where the photoluminescence excitation and emission energies are distributed over a broad range. We demonstrate the performance of the spectrometer by measurements on SWCNTs and we find that it operates with the designed sensitivity.

\begin{acknowledgments}
Work supported by the ERC-259374 (Sylo) and the FP7-GA-607491 (COMIQ) grants.
\end{acknowledgments}


\end{document}